\documentclass[epj]{svjour}
\usepackage[latin1]{inputenc}
\usepackage{graphicx}
\usepackage{graphics}
\usepackage{amsfonts}
\usepackage{color}
\usepackage[T1]{fontenc}
\usepackage[english]{babel}
\usepackage{wrapfig}
\usepackage{amsmath}
\usepackage{here}
\usepackage{amssymb}
\usepackage{mathrsfs}
\usepackage[final]{pdfpages}
\usepackage{graphics}
\begin{document}
\title{Landau levels, self-adjoint extensions and Hall conductivity on a cone}
\author{A. Poux\inst{1} \thanks{\emph{Present address:UFR Facult\'{e} de Sciences et d\'{}Ing\'{e}nierie , Universit\'{e} Toulouse III - Paul Sabatier, Toulouse, 31062 cedex 9,  France}}\and L. R. S. Ara\'{u}jo\inst{1}\and C. Filgueiras\inst{1} \and F. Moraes\inst{2}
%
}                     
%
%
\institute{Unidade Acad\^{e}mica de F\'{i}sica, Universidade Federal
de Campina Grande, POB 10071 Campina Grande, PB 58109-970, Brazil \and Departamento de F\'{i}sica, Universidade Federal da Para\'{\i}ba, Caixa Postal 5008, Jo\~ao Pessoa, PB 58.051-900, Brazil}
\date{Received: date / Revised version: date}
%
\abstract{
In this work we obtain the Landau levels and the Hall conductivity at zero temperature of a two-dimensional electron gas on a conical surface. We investigate the integer quantum Hall effect considering two different approaches. The first one is  an extrinsic approach which employs an effective scalar potential that contains both the Gaussian and the mean curvature of the surface. The second one,  an intrinsic approach where the Gaussian curvature is the sole term in the scalar curvature potential.  From a theoretical point of view, the singular Gaussian curvature of the cone may affect the wave functions and the respective Landau levels. Since this problem requests {\it self-adjoint extensions}, we investigate how the conical tip could influence the integer quantum Hall effect, comparing with the case were the coupling between the wave functions and the conical tip is ignored. This last case corresponds to the so-called {\it Friedrichs extension}. In all cases, the Hall conductivity is enhanced by the conical geometry depending on the opening angle. There are a considerable number of theoretical papers concerned with the self-adjoint extensions on a cone and now we hope the work addressed here  inspires experimental investigation on these questions about quantum dynamics on a cone. 
\PACS{
      {PACS-key}{discribing text of that key}   \and
      {PACS-key}{discribing text of that key}
     } 
} 
\maketitle
%
\section{Introduction}
The Physics of the non-planar two-dimensional electron gas has become an active area of research \cite{magaril}. Curvature induces novel phenomena which might be important to engineering quantum devices \cite{saxena}. On the other hand, curvature can influence  well known two-dimensional phenomena. For example, in references \cite{ferrari1}, the cylindrical geometry has profound effects in a two-dimensional electron gas (2DEG) in an external magnetic field. Hall conductance in a spherical cap was investigated in \cite{cap}. Landau levels and the Hall conductivity were also investigated theoretically on a surface of constant negative curvature in \cite{bulaev}. The magnetic moment of a 2DEG in this geometry was addressed in \cite{bula}. A free particle moving on a one-sheeted hyperboloid was discussed both at the classical and quantum levels and novel equations of motion were found in \cite{k2}. Quantization on a prolate and on a oblate ellipsoid, considering applications to multiple-shell fullerenes, can be found in \cite{ivailo}. A special surface on which many theoretical investigations are realized is the cone. Its geometry is associated to  topological line defects in semiconductors \cite{katanaev} known as disclinations  . We can cite some examples as the investigation of the influence of this surface on the specific heat\cite{heat} and on the persistent currents \cite{sergio} of an electron gas.  A particle constrained to move on a cone and bound to its tip by harmonic oscillator and Coulomb-Kepler potentials was considered in \cite{adam}. The quantum mechanics on a cone was studied by the path integral method in \cite{inomata}. In \cite{janaira}, the Landau levels on conical graphene were obtained. An interesting problem, concerned with the possibility of holonomic quantum computation based on the defect-mediated properties of graphite cones, was investigated in \cite{knut}. 

A fundamental problem that has been discussed in \cite{cone1} is how the cone tip affects the quantum dynamics on a cone. From a mathematical point of view, there  exists a coupling between the wave functions and the singular scalar curvature on a cone \cite{remarks}. However, real cones have smoothed out tips and therefore no singularity. The influence of such  coupling in real systems can only be  investigated from an experimental point of view. Considering this, we investigate the  Landau levels on a cone  in both cases: first without  taking into account the influence of the singularity and then using the self-adjoint extension, which is the mathematically correct way to incorporate it. We then compare both cases and describe how should  be the Integer Quantum Hall effect in both scenarios, hopping that these analysis would inspire experimental investigations of singular effects on $ \rm 2DEG^{'}s$. Although our findings are applied only to common semiconductors, they give a general idea on how singularities can affect a 2DEG in other materials. For instance, we expect the work addressed may lead to further theoretical and experimental investigations in graphene sheets  \cite{graphene} as well as in 2D topological insulators \cite{top} with conical tips.

We will see that the profile of the Hall conductivity versus the external magnetic field changes whether or not we have  a coupling with the conical tip.  We also investigate the integer quantum Hall effect using two different models for electrons on a curved surface. One is based on the dimensional reduction of squared Dirac theory on surfaces employing an intrinsic approach which implies an effective scalar potential proportional to the Gaussian curvature on the surface in the non relativistic limit. The other employs an extrinsic approach which implies an effective scalar potential which contains both the Gaussian and the mean curvature on the surface. This discrepancy was addressed in \cite{remarks}. We will show that the profile of the Hall conductivity may show discrepancies as well. Depending on the opening angle of the cone, the contribution from the mean curvature potential may be irrelevant. 

This work is divided as follows:  in the section 2, we obtain the Landau levels on a cone and we discuss how the conical tip affect them. In section 3, we compute the Hall conductivity at zero temperature in order to show how the presence/absence of singular effects changes its profile. We also investigate how the results change depending on which theoretical model for carries on curved surfaces is considered. In the last section we have the concluding remarks.

\section {Landau levels for a particle on a cone}

Let us consider the coordinates $l$ and $\varphi$ of a particle confined to the surface of a cone  as defined by
\begin{equation}
\left\lbrace
\begin{array}{lll}
$   $x=l\sin\alpha\cos\varphi\\
		y=l\sin\alpha\sin\varphi\\
		z=l\cos\alpha$
$		\end{array}
		\right.\;,\label{cone}
\end{equation}
where $0\leq\varphi\leq 2\pi$ is the usual cylindrical $\varphi$ coordinate and $0< l <+\infty$ (see Fig. 1). Note that $l\sin\alpha$ is the usual cylindrical $\rho$  coordinate.
Consider an applied uniform magnetic field in the $z-$direction, $\vec{B}=(0,0,B_{z})$, as depicted in Fig. 1. For this field, we may choose  the vector potential $\vec{A}=\frac{1}{2}\rho B_z \vec{\hat{\varphi}}$ such that $\vec{B}=\vec{\nabla}\times \vec{A}=B_z \hat{\vec{z}}$. Then, 
\begin{equation}
\vec{A}=\frac{B_{z}l}{2}\sin\alpha \vec{\hat{\varphi}}\;.
\end{equation}

\begin{figure}[!htb]
\begin{center}
\includegraphics[height=7cm]{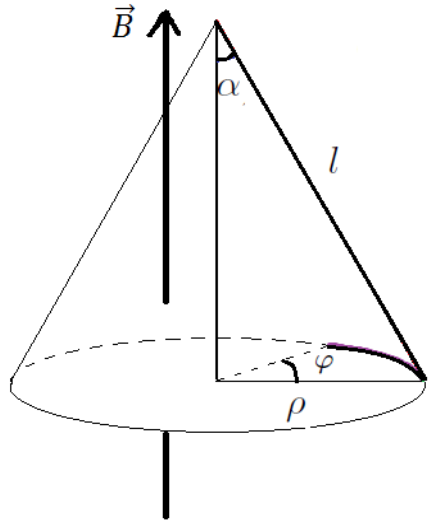}
\caption{Geometrical setting of the problem:   section of height $l\cos\alpha$ of the infinite  straight circular cone of opening angle $2\alpha$ and uniform magnetic field parallel to the cone axis. We consider $\alpha$ between $0$ and $\pi/2$, since, for $\pi/2<\alpha <\pi$ we just have an inverted cone.}
\label{cone}
\end{center}
\end{figure}

We can now write the Schr\"odinger equation for a charged particle confined  to the surface of a cone immersed in an external magnetic
 field in the $z-$direction. From the relation (\ref{cone}), the map of a cone is $\vec{X}(l,\varphi)=(l\sin\alpha\cos\varphi,l\sin\alpha\sin\varphi,l\cos\alpha)$.  The coefficients of the induced metric are given by  \cite{docarmo}
\begin{equation}
g_{11}=X_l\cdot X_l=\frac{\partial \vec{X}}{\partial l}\cdot \frac{\partial \vec{X}}{\partial l}=1\;,
\end{equation}
\begin{equation}
g_{21}=g_{12}=X_l\cdot X_\varphi=\frac{\partial \vec{X}}{\partial l}\cdot \frac{\partial \vec{X}}{\partial \varphi}=0\;,
\end{equation}
\begin{equation}
g_{22}=X_\varphi\cdot X_\varphi=\frac{\partial \vec{X}}{\partial \varphi}\cdot \frac{\partial \vec{X}}{\partial \varphi}=l^2\sin^2\alpha\;.
\end{equation}
 So, the metric tensor and its inverse are
\begin{equation}
g_{ij}=
\begin{pmatrix} 
1&0\\ 
0&l^{2}\sin^{2}(\alpha)\\ 
\end{pmatrix} 
\end{equation}
and
\begin{equation}
g^{ij}=
\begin{pmatrix} 
1&0\\ 
0&\frac{1}{l^{2}\sin^{2}(\alpha)}\\ 
\end{pmatrix} \;,
\end{equation}
respectively.
We consider the usual minimal coupling Hamiltonian for a charged particle 
\begin{equation}
\hat{H}=\frac{1}{2m}(\hat{\vec{p}}-e\hat{\vec{A}})^2\;.
\end{equation}
In order to ensure the Hermiticity of the radial momentum operator, $\hat{p}_l$, acting on the radial wave functions must have the form \cite{acsy}
\begin{equation}
\hat{p}_l\psi(l)=-i\hbar\left(\frac{\partial}{\partial l}+\frac{1}{2l}\right)\psi(l)\;\label{herm}
\end{equation}
whereas \cite{cone1}
\begin{equation}
p_{\varphi}=-\frac{i\hbar}{l\sin\alpha }\partial_{\varphi}.
\end{equation}
So, we achieve 
\begin{equation}
\hat{H}=-\frac{\hbar^2}{2m}\left[\frac{1}{l}\frac{\partial}{\partial_l}\left(l\frac{\partial}{\partial_l}\right)-\frac{1}{4l^2}-\left(-\frac{i}{l\sin\alpha}\frac{\partial}{\partial_\varphi}-\frac{e\sin\alpha B_{z}l}{2\hbar}\right)^2\right]
\end{equation}
for the particle on  a cone under the action of a magnetic field. It is important to notice that the Schr\"odinger equation for a  particle moving on curved surfaces exhibits a scalar potential which describes its interaction  with geometry. It has been shown by Ferrari and Cuoghi {\cite{ferrari} that this potential does not couple with the magnetic field. The geometric interaction  is given by the potential
\begin{equation}
V_S=-\frac{\hbar^2}{2m}\left({\rm M}^2-{\rm K}\right)=-\frac{\hbar^2}{2m}\left(k_1-k_2\right)^2\;,\label{dacosta}
\end{equation}
where, $k_1$ and $k_2$ are the surface's {\it principal curvatures}, ${\rm M\equiv1/2(k_1+k_2)}$ is the {\it mean curvature} and $K\equiv k_1k_2$ is the {\it Gaussian curvature} of the  surface \cite{docarmo}. This potential was first derived in \cite{dacosta}. Curiously, in Geometry, the term $W=\int_{S} \left(k_1-k_2\right)^2 dA$ is known as the Willmore energy \cite{Willmore} which, remarkably, is invariant under conformal transformations of $\mathbb{R}^3$ \cite{White}. The Willmore energy gives a measure of how much a given compact surface deviates from a sphere (which has $W=0$). It has appeared naturally in other physical contexts, like for example, the Helfrich model \cite{Helfrich} for elasticity of cell membranes. 

Finally, the Schr\"odinger equation is $\left[\hat{H}+V_S\right]\Psi(l,\varphi)=E\Psi(l,\varphi)$. So, we have
\begin{equation}
-\frac{\hbar^2}{2m}\left[\frac{1}{l}\frac{\partial}{\partial l}\left(l\frac{\partial}{\partial l}\right)-\frac{1}{4l^2}-\left(\frac{-i}{l\sin\alpha}\frac{\partial}{\partial_\varphi}-\frac{e\sin\alpha B_{z}l}{2\hbar}\right)^2\right]\Psi(l,\varphi)+V_S\Psi(l,\varphi)=E\Psi(l,\varphi)\label{scho}\;.
\end{equation}
We may calculate $\rm M$ and $\rm K$ by the formula \cite{docarmo}
\begin{equation}
{\rm M}=\frac{1}{2}\frac{eG-2fE+gE}{EG-F^2}\;
\end{equation}
and
\begin{equation}
{\rm K}=\frac{1}{2}\frac{eg-f^2}{EG-F^2}\;,
\end{equation}
respectively. For a cone, $E=g_{11}=1$, $F=g_{12}=0$, $G=g_{22}=l^2\sin^2\alpha$, $e=\vec{n}\cdot \frac{\partial^2\vec{X}}{\partial\varphi^2}=0$, $f=\vec{n}\cdot \frac{\partial^2\vec{X}}{\partial\varphi\partial l}=0$ and $g=\vec{n}\cdot\frac{\partial^2\vec{X}}{\partial l^2}=l\sin\alpha\cos\alpha$.  This way, ${\rm M}=\frac{\cos\alpha}{2\sin\alpha l}$ and $K=0$ except at the tip of the cone where it is a delta-function singularity.  We consider first the case of  real cones, which do not have singularities, we neglect the influence the effect of the singular Gaussian curvature on the wave function.  By considering the wavefunction as $\Psi(l,\varphi)=\Psi(l)e^{ij\varphi}$,  with $j=0,\pm1,\pm2,...$, we arrive at 
\begin{equation}
-\frac{\hbar^2}{2m}\left[\frac{1}{l}\frac{d}{d l}\left(l\frac{d}{d l}\right)-\frac{\frac{1}{4}+\frac{\mu^2}{\sin^2\alpha}}{l^2}\right]\Psi(l)+\frac{mw^2l^2}{2}\Psi(l)=\Sigma\Psi(l)\label{harm}
\end{equation}
where $\mu^2=j^2-\frac{1-\sin^2\alpha}{4}$, $\omega=\left(\omega_c/2 \right)\sin\alpha$, with  $\omega_c=\frac{eB_z}{m}$(cyclotron frequency), and
\begin{equation}
\Sigma=E+j\hbar\omega\;.\label{cond}
\end{equation}
Notice that, in general, $\frac{1}{4}+\frac{\mu^2}{\sin ^2 \alpha}>0$ but for the specific case of $j=0$ and $0<\alpha<\pi/4$ it is negative.   This means that we have an attractive $1/l^2$ potential near the cone tip. This is a pathological potential  whose bound state spectrum is unbound from below \cite{case} and should be treated with proper regularization. As it will be seen below, in order for the wave function to be square integrable we must have $j=0$ and then our study, for simplicity, will be restrict to the cones with $\alpha$ in the interval $[\pi/4,\pi/2]$.

 The differential equation (\ref{harm}) describes a two-dimensional quantum oscillator on a conical background. This problem was addressed in \cite{kowaslki} for a circular double cone. In the case of 2DEG on a single cone,  the range of the coordinate $l$ is $0< l <+\infty$ while that for a circular double cone it is $-\infty< l <+\infty$ . Like in \cite{kowaslki},  the wavefunctions are  given by
\begin{equation}
\Psi(l)={\rm C} \left|l\right|^s \exp(-\frac{m\omega^2 l^2}{2})U\left(\frac{s}{2}+\frac{1}{4}-\frac{E}{2\omega},s+\frac{1}{2},m\omega l^2\right)\label{wf}
\end{equation}
 with 
\begin{equation}
s=\frac{1}{2}\left(1\pm\sqrt{1+\frac{4\mu^2}{\sin^2\alpha}}\right)\label{ss}
\end{equation}
and {\rm C} being the normalization constant. $U(a,b,c)$ is the Kummer function \cite{abra}. In order to have proper normalization of the wavefunction,we must have $$\lim_{l\rightarrow\infty}\Psi(l)\rightarrow0\;.$$ In order to get this condition, the series $U\left(\frac{s}{2}+\frac{1}{4}-\frac{E}{2\omega},s+\frac{1}{2},m\omega l^2\right)$ in (\ref{wf}) must be a polynomial of degree n. This is achieved when \cite{abra}
\begin{equation}
\frac{s}{2}+\frac{1}{4}-\frac{E}{2\omega}=-n\;.
\end{equation}
No extra condition must be imposed on $\Psi$. On the other hand, we have to investigate the behavior of the density of probability as $l\rightarrow0$, that is, 
\begin{equation}
\lim_{l\rightarrow0}\left|\Psi(l)\right|^2 ldl\rightarrow0\;.
\end{equation}
Considering \cite{abra} $$\lim_{x\rightarrow0}U(a,b,x)\rightarrow \frac{\pi}{\sin(b\pi)}\left[\frac{1}{\Gamma(1+a-b)\Gamma(b)}-\frac{x^{1-b}}{\Gamma(a)\Gamma(2-b)}\right]\;,$$
we get
\begin{equation}
\lim_{l\rightarrow0}\left|\Psi(l)\right|^2 ldl=C l^{2s+1}\;,\label{si}
\end{equation}
where $C$ is a constant. Before doing any further analysis, we must mention that, as it was argued in \cite{cone1,kowaslki}, the tip of the cone correspond to a singularity and the the quantum dynamics requests {\it self-adjoint extensions} \cite{fr,selfadj}, in contrast with the double cone case, where such singular effects do not show up. Notice that the case of quantum particles around a double cone is similar to the one where we have quantum particles on a single cone without the coupling between the wavefunctions and a singular curvature. This is a situation without $\delta$-function potentials, which correspond the the so-called {\it Friedrichs extension} \cite{fr}. It is characterized by the following boundary condition,
\begin{equation}
\lim_{l\rightarrow0}l\partial_l \psi(l)=0\;.
\end{equation}
This condition means that, although we have a conical tip, the wavefunctions are {\it regular} at $l=0$. When we consider the $\delta-$function potential, the quantum dynamics requests self-adjoint extensions since the wavefunctions now are {\it irregular} as \cite{eu} $l\rightarrow0$. 
In what follows, we evoke the results of \cite{kowaslki} for the regular case and after that we discuss what is going to change when we take into account the wavefunctions being irregular at the conical tip.

For electrons on a single cone without the coupling with the singular Gaussian curvature, the parameter $s$ above must be just that with a plus sign. This statement comes from equation (\ref{si}). Notice that, for positive $s$, $\left|\Psi(l)\right|^2 ldl$ goes to zero as $l\rightarrow0$.  Since we do not have singularity, this means that the wavefunctions are regular at the conical tip. So, the eigenvalues are given by
\begin{equation}
\Sigma_{j,n}=2\hbar\omega\left(n+\frac{1}{2}+\frac{1}{4}\sqrt{1+\frac{4\mu^2}{\sin^2\alpha}}\right).
\end{equation}
Combining this equation with (\ref{cond}), we arrive at the {\it Landau levels } for electrons on a cone, that is, 
\begin{equation}
E_{j,n}=\hbar\omega_c\sin\alpha\left(n+\frac{1}{2}+\frac{1}{4}\sqrt{1+\frac{4\mu^2}{\sin^2\alpha}}-\frac{j}{2}\right).\label{ll}
\end{equation}
The problem addressed here has an important difference from that investigated in \cite{kowaslki}. First of all, (\ref{ll}) has an extra piece proportional to $-j/2$. Also, the term inside the square root in (\ref{ll}) shows the parameter $\mu^2$( instead of just $j^2$) which can never be null for integer values of $j$. If we take $j=0$ in equation (\ref{harm}) we get the same result as taking $j\rightarrow0$ in  (\ref{ll}), in contrast with \cite{kowaslki}, where the results are different. This solution corresponds to the case of an harmonic oscillator on a meridian of the cone. This difference comes from the geometric potential $V_S$ which leads to an inverse square distance dependent quantum potential. 

We now turn our attention to the case where there is a coupling between the wavefunctions and the singular Gaussian curvature. The problem of the quantum harmonic oscillator with singularities was addressed in \cite{eu}.  Everything we have done so far remains the same but the parameter $s$ (\ref{ss}) now admit both signs, plus and minus. The case with minus sign correspond to wave solutions which are irregular at the tip of the cone. In order to have $\int\left|\Psi(l)\right|^2 ldl=C l^{2s+2}\rightarrow0$ as $l\rightarrow0$, we must impose the following constrain
\begin{equation}
-1<s<1\;.
\label{cond1}
\end{equation}
 This constraint guarantees that the wavefunctions are square integrable. Whenever we have singularity, this condition must happens \cite{hagen}. In order to fulfill (\ref{cond1}), the only possible value for the angular momentum is $j=0$.  Then, the Landau levels split as 
\begin{equation}
E_{0,n}^+=\hbar\omega_c\sin\alpha\left(n+\frac{1}{2}+\frac{1}{4}\sqrt{2-\frac{1}{\sin^2\alpha}}\right)\label{ll1}
\end{equation}
and
\begin{equation}
E_{0,n}^-=\hbar\omega_c\sin\alpha\left(n+\frac{1}{2}-\frac{1}{4}\sqrt{2-\frac{1}{\sin^2\alpha}}\right)\;.\label{ll2}
\end{equation}

In summary, without singular effects, the Landau levels are those obtained as (\ref{ll}), with $j=0,\pm1,\pm2,...$. If the singular Gaussian curvature plays a role, then the only possibility will be $j=0$. In the former case, the degeneracy is infinitely broken in comparison to the flat case($j=0,\pm1,\pm2,...$ ) and in the latter one, each Landau level is split twice. We sketch the Landau levels for both problems in figure \ref{eg}. In both cases, each energy labeled with index $n$ splits in two energy values but they show different energy spacing. This will have some impact in the Hall conductivity profile.

We note that the energy of the Landau levels as expressed by (\ref{ll1}) and (\ref{ll2}) becomes complex when the cone angle $\alpha$ is not in the interval $[\pi/4,\pi/2]$. As pointed out above, for cones with such values for $\alpha$, a pathological contribution to the potential appears and the problem requires special regularization at $l=0$. For these cones the result given by expressions (\ref{ll1}) and (\ref{ll2}) is therefore not valid. Since we are only considering the cones with $\alpha$  in the interval $[\pi/4,\pi/2]$, the energy levels given by  (\ref{ll1}) and (\ref{ll2}) are always real.

\begin{figure}[!htb]
\begin{center}
a\includegraphics[height=4cm]{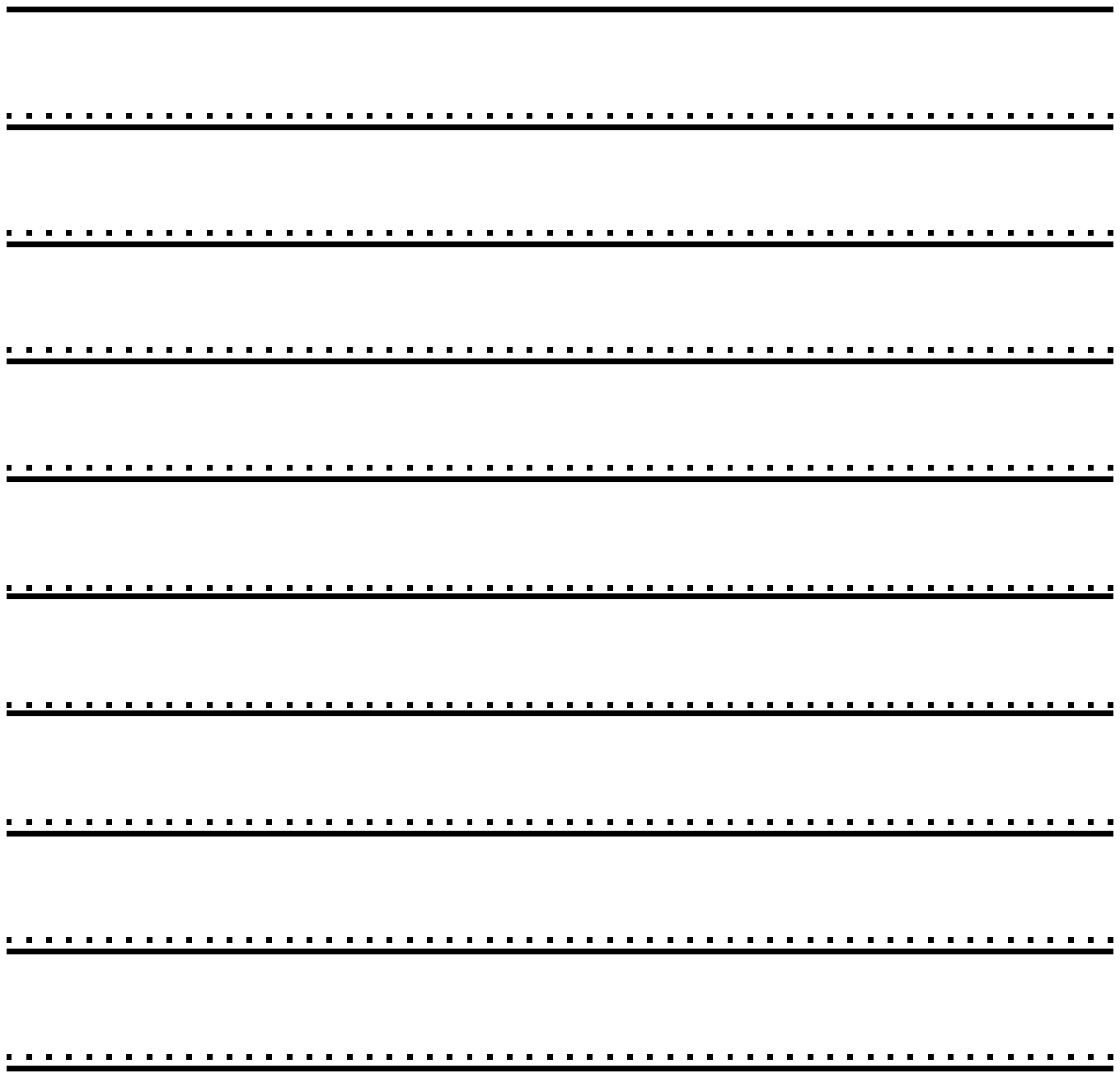}\label{f1}
			\hspace{0.2cm}
b\includegraphics[height=4cm]{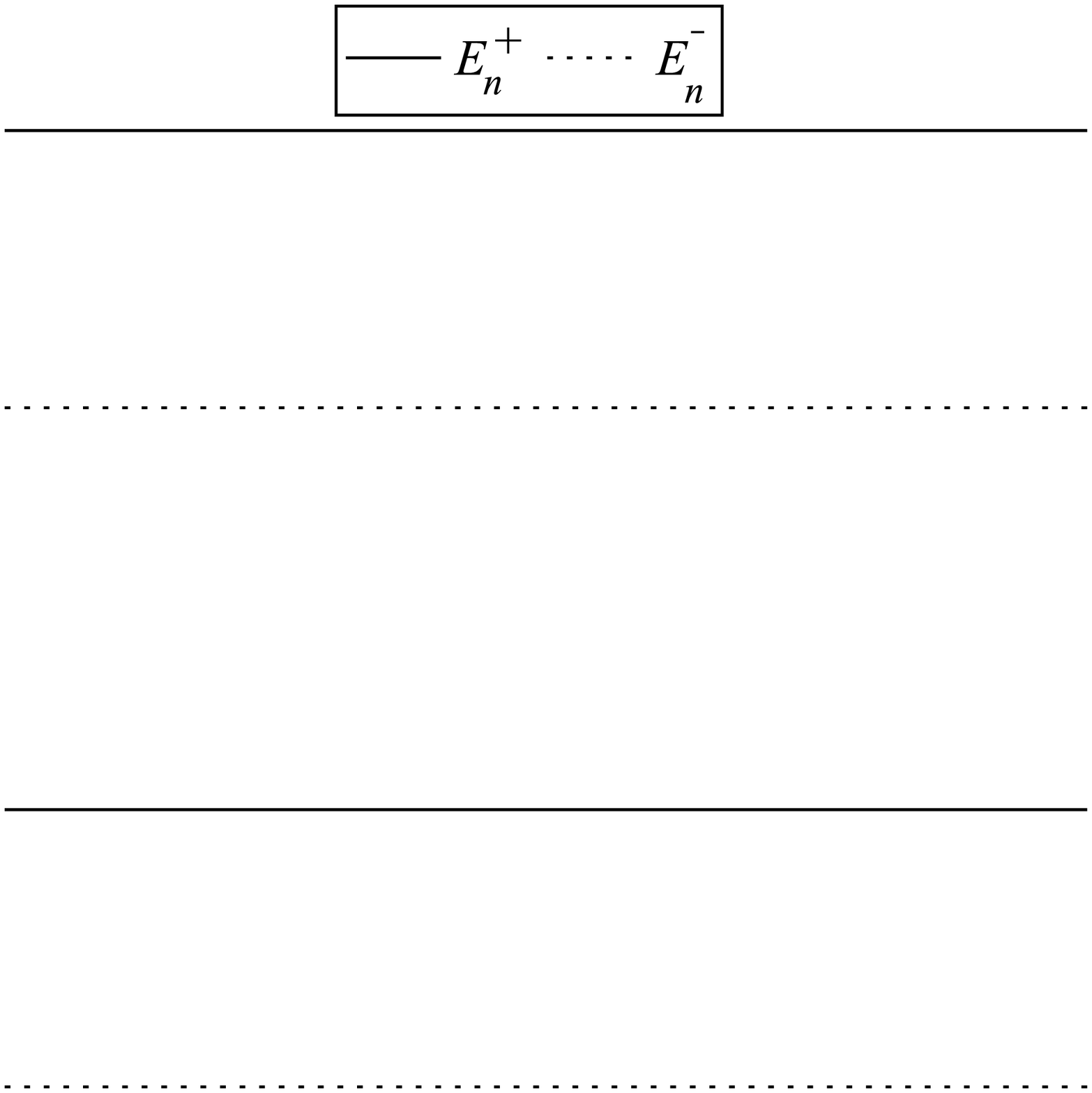 }\label{f2}
\caption{{\bf a} Some Landau levels for a 2DEG on a single cone without singular effects(we considered just three filled angular momentum states, $j=-1,0,1$) and {\bf b} with singular effects.  In both cases, each energy labeled with index $n$ splits in two energy values but they show different energy spacing. We considered $\alpha=\pi/3$ rad.}\label{eg}
\end{center}
\end{figure}

With respect to the Landau levels given by (\ref{ll1}) and (\ref{ll2}) we remark that, for $\alpha=\pi/4$, the levels become degenerate. This is probably related to some specific symmetry of the $\alpha=\pi/4$ cone. A possible explanation might come from Supersymmetric Quantum Mechanics (SUSYQM) \cite{cooper} which relates pairs of Hamiltonians which share a particular mathematical relationship (the original SUSY concept related bosons and fermions). A hint for the specific puzzle of this degeneracy might be in the hidden SUSY found in the delta potential problem \cite{correa1} and in the Aharonov-Bohm system \cite{correa2}. This point deserves further investigation as well as the inclusion of the spin degree of freedom which, in the field of a magnetic vortex, shows additional SUSY properties\cite{correa3}.

\section{The Hall conductivity}

In this section, we compute the Hall conductivity of a 2DEG on a cone. The system which we are interested in is essentially the one depicted in Fig. 1 with the electrical contacts missing. It consists in a conical surface in the presence of a uniform magnetic field parallel to its axis. This, in principle, can be achieved with a carbon nanocone or with semiconductors heterostructures grown as a cone. We consider the integer quantum Hall effect, which consists of a quantization of the conductivity and appearance of plateaus at particular values of the external magnetic field. It manifests at low temperatures. We take into account just the ground state, $T=0$. Each Landau level contributes one quantum of conductivity to the electronic transport. If $n_0$ Landau levels are completely filled, the Hall conductivity, $\sigma_H$, will be given by\cite{hall}
\begin{equation}
\sigma_H=-n_0\sigma_o\;,
\end{equation}
where $\sigma_o\equiv e^2/h$ is the quantum of conductivity, $e$ is the electron charge and $h$ is the Planck$^{,}$s constant.

We start the calculations determining $n_0$ for a specific value of magnetic field. This is achieved by counting how many states are occupied below the {\it Fermi energy}, $E_F$. We compare $E_F$ with the Landau levels showed in figures \ref{eg}, making it equal to the highest Landau level filled.  Next, we determine the quantity of magnetic field required to move the Hall conductivity from  $-n_0\sigma_o$ to $-(n_o-1)\sigma_o$. In doing that, we are, in fact, determining the size of the plateaus at zero temperature for a specific $n_o$. Continuing this procedure, we can plot the Hall conductivity versus the plateaus. In our calculations, we start at $B=10T$.

In figure \ref{alpha} we plot the Hall conductivity for three different cases: a flat sample without singularity and the cone with and without singularity. In the last case we considered the filled states $j=-1,0,1$ only. If we take more values of $j$, which is not allowed for the Landau levels in the singularity case, we will obtain other intermediate plateaus. We see that the plateaus of conductivity on the cone are shifted to higher magnetic fields and there is an enhancement of each Hall conductivity in both cases. We can see clearly that the singular boundary conditions change significantly the Hall conductivity profile.  We also probe the influence of the opening angle $\alpha$. In the interval of magnetic field showed, the Hall conductivity profile crosses the one  for a flat sample for some values of $\alpha$ and some plateaus are shifted to lower magnetic fields. We can also observe that, for $\alpha=\frac{\pi}{2}$, we do not reach what we are calling {\it flat case}. For it, we considered just the Landau levels $E_n=\hbar\omega_c(n+1/2)$. We can not recover it here because of equation (\ref{herm}). This makes sense since we can not make any transformation with the conical tip which is a singular point. We believe that the case $\alpha=\pi/2$ correspond to a flat sample with an extracted point. The self-adjoint extensions in this case was addressed in \cite{extra} and we think it could be probed experimentally based on the Hall effect as discussed here. With these plots, we can more easily see that we have a reduction of the size of the plateaus in comparison to the flat case. In analyzing figure \ref{alpha}, we observe that considering or not the singular effects on the wavefunctions is an important issue on a theoretical basis. We hope this fact may be experimentally probed.

\begin{figure}[!htb]
\begin{center}
(a)\includegraphics[height=7cm]{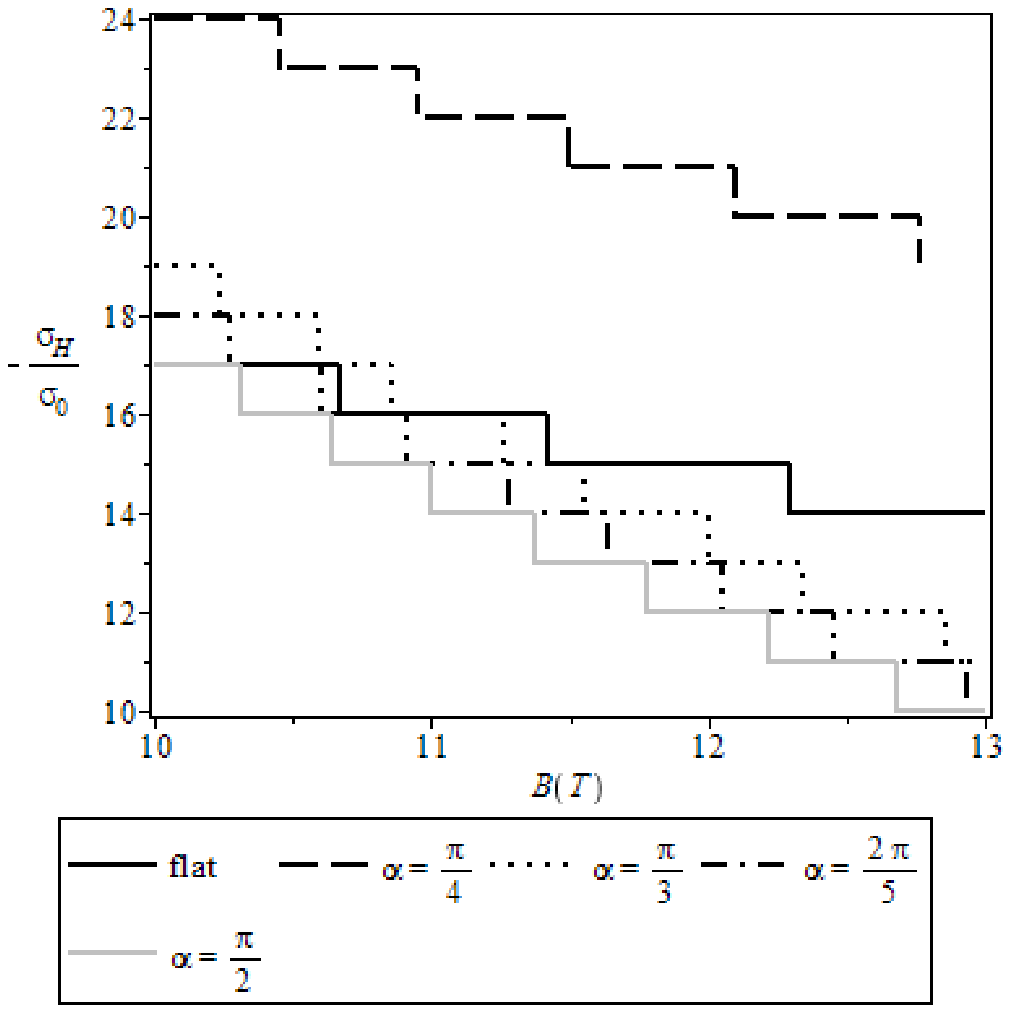}
			\hspace{0.5cm}
(b)\includegraphics[height=7cm]{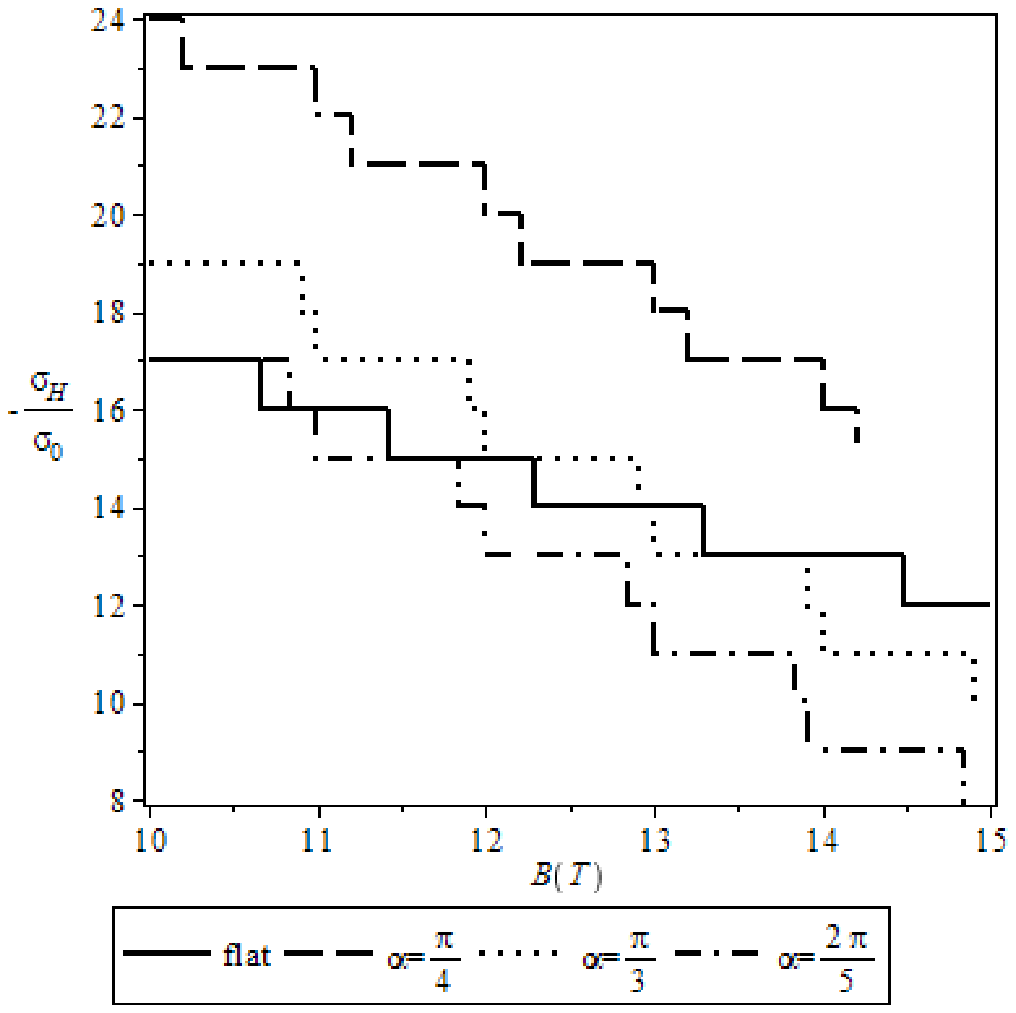}
	\hspace{0.5cm}
\caption{Hall conductivity on the cone for different opening angles. {\bf (a)} irregular and {\bf (b)} regular case. There is an enhancement on the Hall conductivity as $\alpha$ is reduced. Notice that considering the singularity in the wavefunctions has important consequences.}
\label{alpha}
\end{center}
\end{figure}

We now move to see the influence of the mean curvature, that is, we probe the two models we are dealing with for electrons on a curved surface. We consider both cases, with and without singular effects. In the figure \ref{M}, we can see that, depending on the opening angle $\alpha$, the two models may yield or not different quantitative results. 
Indeed, for the singular case and for $\alpha=\frac{\pi}{3}$, the plateaus are shifted to lower magnetic fields. This happens because of the size of the plateaus, which are smaller when we consider the mean curvature potential. For $\alpha=\frac{2\pi}{5}$, taking  or not into account the mean curvature potential does not make any difference.
Without singular effects, the same thing happens. The difference is only that pointed out in figures \ref{alpha}.

\begin{figure}[!htb]
\begin{center}
(a)\includegraphics[height=6cm]{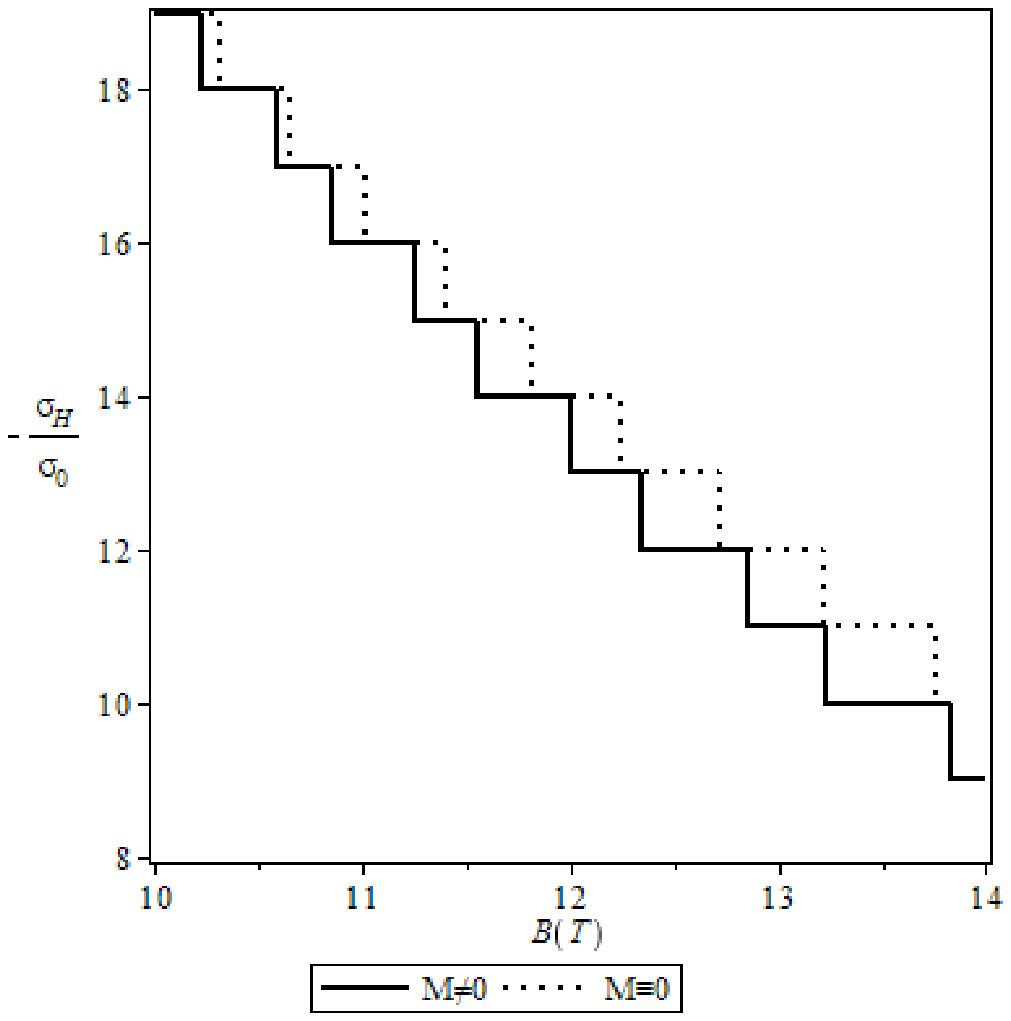}
			\hspace{0.5cm}
(b)\includegraphics[height=6cm]{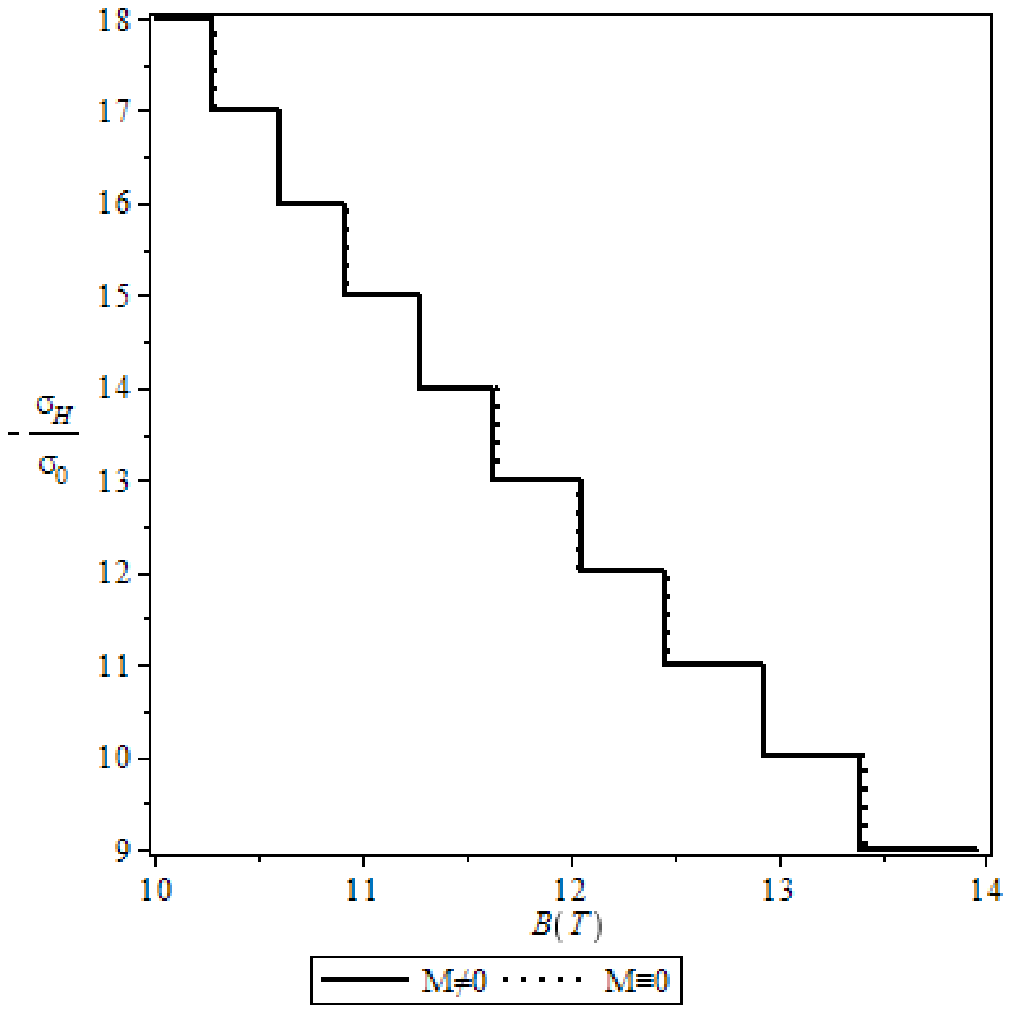}
	\hspace{1.0cm}
(c)\includegraphics[height=6cm]{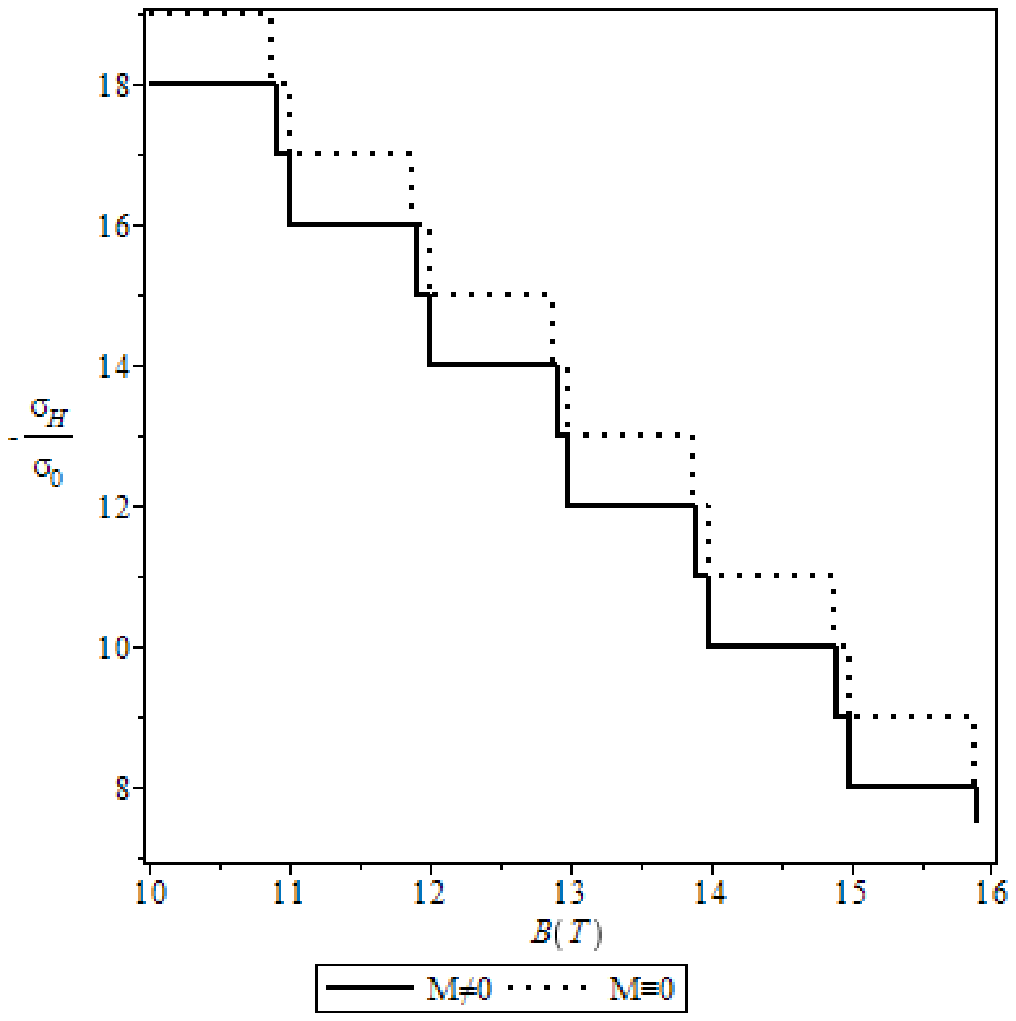}
	\hspace{0.5cm}
(d)\includegraphics[height=6cm]{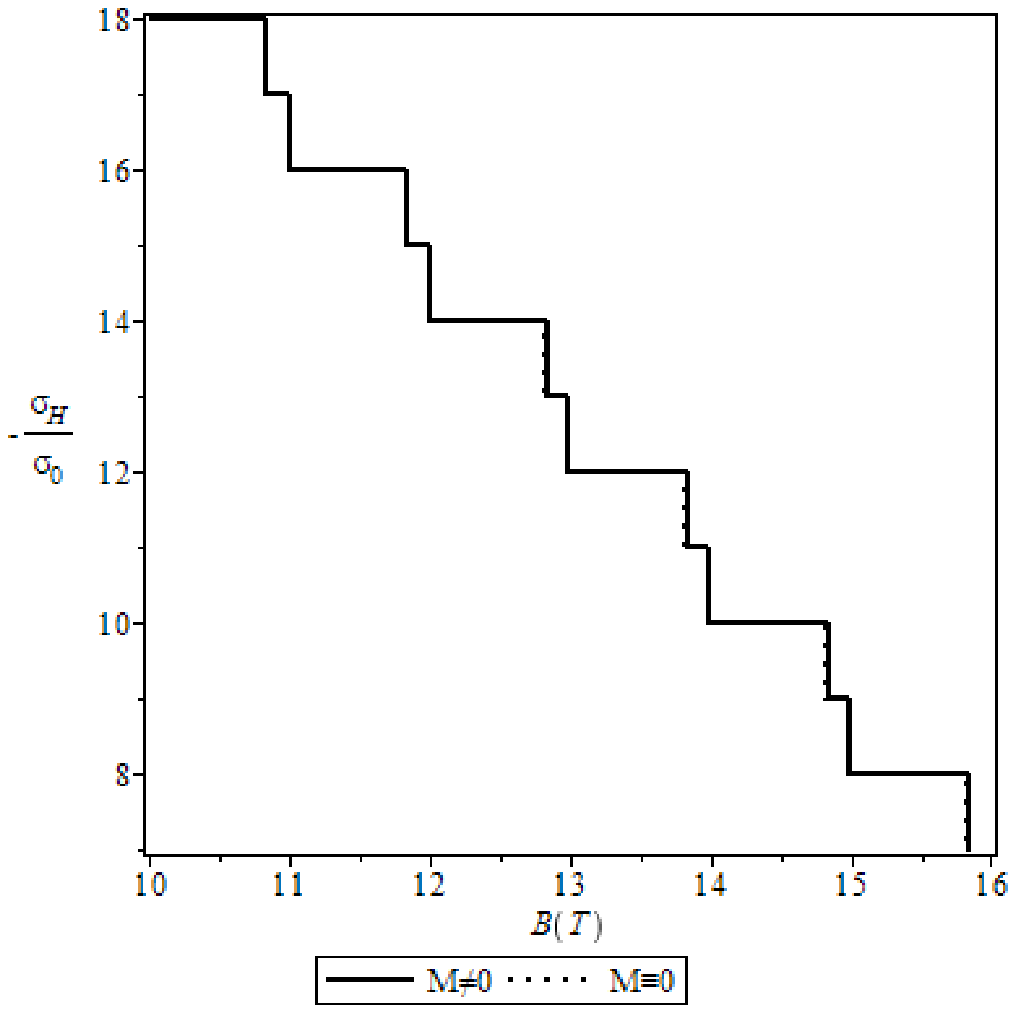}
\caption{Hall conductivity on the cone with and without the mean curvature potential ${\rm M}$:  without singularity, we have {\bf (a)}  $\alpha=\frac{\pi}{3}$ and {\bf (b)} $\alpha=\frac{2\pi}{5}$, while with singularity we have {\bf (c)}  $\alpha=\frac{\pi}{3}$ and {\bf (d)} $\alpha=\frac{2\pi}{5}$. The difference between both approaches may be noted depending on the opening angle $\alpha$ considered(see figure \ref{fonction}).}
\label{M}
\end{center}
\end{figure}

Plotting the mean curvature potential $M$ in function of the opening angle $\alpha$, we obtain figure \ref{fonction}. That explains clearly what happens in figure \ref{M}. Indeed, when we have a big opening angle, we can see that the mean curvature potential goes to zero. This is why we have seen in our plots that when we take a big opening angle, taking or not into account $M$ do not make a big difference, contrary to a small opening angle. This said, we conclude that depending on the opening of the cone, the difference between the two approaches discussed here for electrons on curved surfaces will not show up. This fact may happens for other geometries.  So, comparing the mean curvature with the Gaussian curvature first can tell us when it is important to take it into account in our calculations.
Obviously, for {\it minimal surfaces} ($M\equiv0$), the two approaches yield identical Hall profiles.
\begin{figure}[!htb]
\begin{center}
\includegraphics[height=7cm]{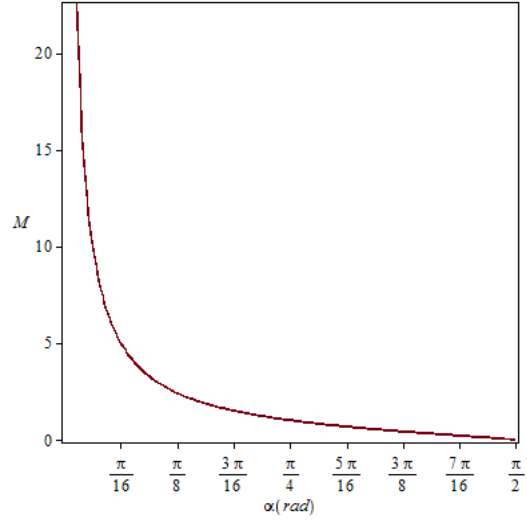}
\caption{Profile of the mean curvature $M$ depending on the opening angle $\alpha$. We can see that, for a smaller values of $\alpha$,  $M$ grows while for higher values of $\alpha$ approaching $\pi/2$, $M\longrightarrow 0$.}
\label{fonction}
\end{center}
\end{figure}
\newpage
\section{Conclusion}
In this work we studied the integer quantum Hall effect on a cone. We have showed that considering the coupling between the wavefunctions and the singular scalar curvature (the conical tip),  the profile of the Hall conductivity versus the external magnetic field modifies in a significant way.  We have also considered two different models for electrons on a curved surface. One contains an effective scalar potential proportional to the Gaussian curvature on the surface  while the other one shows an effective scalar potential which contains both the Gaussian and mean curvature on the surface. So, we have seen that the profile of the Hall conductivity show discrepancies when we compare the two theories, depending on the opening angle of the cone. We restricted our study to cones with opening angle $\alpha$ in the interval $[\pi/4,\pi/2]$ which is enough to show the effects of the conical geometry in the Hall conductivity.  As an open problem we leave the case of more acute cone angles, $0<\alpha<\pi/4$, where the boundary condition at $l=0$ has to include some kind of regularization in order to fix the pathology that gives bound states with no lower bound.

The theory based on the da Costa approach(with the presence of mean curvature) is a result of particle confinement and it is the same for electrons and holes. A similar confining procedure
for relativistic carriers in graphene was first addressed in \cite{tun}. From the results we addressed here, we may note that the model one takes to investigate carries on curved graphene is going to be relevant. Without the confinement procedure, which in our case led
 to the mean curvature contribution to the quantum Hall effect, the theory may not fit well the experiments on curved 2DEG. In the case of graphene cones \cite{grapcone}, the conical geometry, which appears due to the presence of topological defects, must be investigated carefully since the singular Gaussian curvature may also affect the electronic transport in this material.  As pointed out in \cite{hagen}, an inadequate choice of the boundary condition for the electronic wavefunction can lead to contradictory physical results. The authors have showed that the irregular solutions at the conical tip must be considered. So, although we do not consider graphene here, our results may corroborate with this statement but we think that it should be probed experimentally in the context of quantum Hall effect.

In summary, the study carried out here can be realized in common semiconductors. However, it shows important features if one has intention to deal with non planar 2DEG on other classes of materials. 

%
%
%

%
%

\end{document}